\documentclass[twocolumn,twoside,showkeys,preprintnumbers,amsmath,amssymb,secnumarabic, nofootinbib, nobibnotes,epl]{revtex4}

\usepackage{epsfig}

\usepackage{graphicx}

\usepackage{fancyhdr}

\usepackage{pslatex}

\newcommand{\sect}[1]{\setcounter{equation}{0}\section{#1}}

\begin{document}



\title{Nuclear problems in astrophysical $q$-plasmas and environments}



\author{M. Coraddu$^{1,\ast}$, M. Lissia$^{1,\dag}$, P. Quarati$^{1,2,\ddag}$, A.M. Scarfone$^{2,3,\S}$}
\affiliation{$^1$Istituto Nazionale di Fisica Nucleare
(INFN), Sezione di Cagliari, I-09042, Monserrato, Italy\\
$^2$Dipartimento di Fisica - Politecnico di Torino,
I-10129, Italy\\
$^3$Istituto Nazionale per la
Fisica della Materia (CNR--INFM), Sezione del Politecnico di Torino,
I-10129, Italy}



\begin{abstract}
Experimental measurements in terrestrial laboratory, space and astrophysical observations of variation and fluctuation of nuclear decay constants, measurements of large enhancements in fusion reaction rate of deuterons implanted in metals and electron capture by nuclei in solar core indicate that these processes depend on the environment where take place and possibly also on the fluctuation of some extensive parameters and eventually on stellar energy production. Electron screening is the first important environment effect.
We need to develop a treatment beyond the Debye-H\"uckel screening approach, commonly adopted within  global thermodynamic equilibrium. Advances in the description of these processes can be obtained by means of $q$-thermostatistics and/or superstatistics for metastable states. This implies to handle without ambiguities the case $q<1$.
\end{abstract}


\maketitle

\thispagestyle{fancy}

\setcounter{page}{1}

\sect{Introduction}

Many physical laws are established in the frame of the hypothesis that each single event of a component of an observed system is independent on any other event. For instance, the nuclear decay law is based on the hypothesis that each nucleus decays without being influenced by the other nuclei and that all nuclei have the same probability of decaying.\\
If $P(dt)\equiv\lambda\,dt$ is the probability of decay of a nucleus in the time interval $dt$, the probability of surviving after $n$ intervals $dt$ is given by the power law
\begin{equation}
(1-P(dt))^n=(1-\lambda\,dt)^n=\left(1-\lambda\,{t\over n}\right)^n \ .
\end{equation}
By making $n$ going to infinity the exponential law $\exp(-\lambda\,t)$ is obtained \cite{Meyerhof,Boscaino,Lissia,Wilk}.\\
Many physical phenomena are studied by assuming that all typical events are independent and non correlated as, for instance, the nuclear reaction events that occur to form nuclear fusion rates in stellar cores.\\ Therefore, one uses the Maxwell-Boltzmann (MB) distribution to describe both the ionic and electronic component. At most, one assumes that all effects from particle correlations and/or non linear effects can be neglected, being their contribution small.\\
Also Debye-H\"uckel (DH) approach to electron screening, developed to take into account the electron and ion influence over the pure Coulomb potential of a given ion charge, is based over  additive, linear effects and standard exponential distributions \cite{Clayton,Rolfs,Johnson,Quarati1}.\\
Very recently, in many different experiments, influence and effects of environment have been observed in addition to the known effects over fusion thermonuclear reactions in stellar cores. One can assume that these effects are  due to correlations, microfield distributions, electron screening, random forces, fluctuations of extensive quantities, among others causes. Here we give a list of the recent observations:
\begin{enumerate}
\item Measurements of electron capture decay rates and constants of $^7$Be in host materials \cite{Huh,Ohtsuki,Hua,Wang,Nir}, of $\alpha$-radioactive uranium exposed to glow-discharge plasma \cite{Dash} and of other heavy nuclei implanted in different environments \cite{Severijns,Jeffesen,Spillane} have recently shown fluctuations and variations of previous standard results.

\item Rates of deuteron-deuteron fusion in metal matrices have shown a great increase respect to the standard laboratory measurements and have shown a non negligible influence of the conducting electrons \cite{Assenbaum,Raiola1,Raiola2,Coraddu,Saltzmann}.

\item Variations observed in electron capture by $^{40}$K could be responsible of observed discrepancies in the density ratio U / Pb in Oklo site phenomenon \cite{Norman1}.

\item Unexpected and unexplained fluctuations in the decay rates of $^{32}$Si and $^{226}$Ra have been reported and evidence of correlations between nuclear decay rates and Earth-Sun distance has been found
    (Jenkins-Fishbach effect \cite{Jenkins1,Jenkins2,http}).\footnote{Very recently, after new measurements, the variation has been reduced and an experimental group suggests that non correlation exists with the Earth-Sun distance, although modulation and variation still exist \cite{Norman2}}.

\item Several careful experiments designed to study the decays of long lived radioactive isotopes have reported observations of small periodic annual variations modulating the well-known exponential decay curve. If Jenkins-Fishbach effect is correct we could have to consider profound consequences for many areas of physics and engineering.
 Discrepancies in half-life determinations of many nuclides reported in literature have been attributed to variations of solar activity during experimental works or to seasonal variations in fundamental constant. However, while in a satellite experiment the correlation between variation of rate and Earth-Sun distance has not been observed \cite{Cooper}, other authors suggest that neutrinos from the flare of 13 December 2006 have induced a change in the decay rate of $^{54}$Mn.

\item Precise half-life of unstable isotopes, in particular of $^{14}$C, are considered in doubt, since the Jenkins-Fishbach effect implies that the decay rate of an isotope is not a constant of Nature, in contrast with the findings of Rutherford, Chadwick and Ellis in 1930 that the rate of transformation of an element is constant under all conditions.
A reasonable interpretation of the so called wiggles in $^{14}$C decay rate has been suggested by Sanders: wiggles indicate that the solar fusion furnace is pulsating in some way like the Cepheyd furnace \cite{Sanders}.

\item Recent work by Sturrock on power spectral analyses of radiochemical solar neutrino data and solar irradiance have revealed modulations attributable to non spherically-symmetric and variable nuclear burning in the solar core \cite{Sturrock}.

\item New determinations of solar CNO content based on 3D hydrodynamic model and a new treatment of departure from Local Thermodynamical Equilibrium (non-LTE) have mined agreement between helioseismology and observations \cite{Scott}.

\item Local non-Gaussianity of the temperature anisotropy of the cosmic microwave background and other non-Gaussianities revealed by satellite experiments have been discussed recently in many papers to which we send for details, since these problems are not object of our interest here.
\end{enumerate}

Of course all the above effects are not concerning with the normal evolution of a star, rather represent fluctuations around average values, the time averages giving rise to constant values. The states at which the nuclear processes take place are not standard global thermodynamic equilibrium states. They can be seen as metastable states with a given life time (eventually a long-lifetime) and therefore the use of $q$-thermostatistics \cite{Tsallis} and/or superstatistics \cite{Wilk1,Beck} can be the appropriate approach to be applied to understand the above effects \cite{Scarfone}.\\
In most of all these processes, the first and most important  environment effect that operates is the electron screening. Usually it is described by DH potential: Coulomb potential which originates from a charge in an astrophysical plasma is screened by the other charges. Decrease of potential is faster than pure Coulomb one because of the exponential factor in DH approximation.\\
To have a more complete description one must go beyond DH potential approach; the electron cloud that screens can be, for instance, more concentrated around the ion that originates the Coulomb potential and its momentum distribution can differ from the Maxwellian one \cite{Quarati1,Coraddu,Quarati2}.\\
Nuclear fusions in stellar cores are influenced by electron screening. Among many nuclear astrophysical processes that are influenced by the surroundings, the electron capture by nuclei, like for instance e$^-+^7$Be$\to^7$Li$+\nu$ in the Sun or e$^-+$p$\to$n$+\nu$ in high density stars, plays a special role because the electron component of the plasma is responsible of the screening and the same electrons are themselves responsible of the screened physical process that is the electron capture by nuclei from bound and continuum states caused by weak nuclear interaction. The first action is due and characterised by the charge distribution in the coordinate space, with this distribution we can build the screening potential. The second action is the weak nuclear process of capture with emission of a neutrino. Its rate is characterised by  the momentum distribution of electrons in the momentum space.\\ Why we are looking for a screening potential different from the widely used DH one?\\If it is required that the distribution in coordinate space be more concentrated around the ion than in DH case, then this can be realized by a $q$-distribution with $q<1$ (depleted tail). For instance, we can show in the case of electron capture by $^8$Be, that the electron distribution tail should be depleted and this fact agrees with the requirement that to obtain an increase of the EC rate (to obtain a decrease of the $^8$B neutrino flux).\\
The number of particles inside the DH sphere is, for many astrophysical systems like stellar core, very low. Therefore standard MB statistics does not apply. The fluctuation of density is large and due to the connection of density with temperature through the equation of state also the fluctuation of temperature is present. We may use superstatistics to justify our choice of non extensive statistics and its distribution.\\
Coulomb cutoff in the spatial space was already used. This is the simplest way to screen the Coulomb field: to cut it at a certain $r=R$ so that the potential is zero outside $R$. No other justifications have been advanced for this approach.\\
A result analogous to this can be obtained considering density and temperature fluctuation in superstatistics or non linear Poisson equation, in such a way that use of the non extensive Tsallis statistics with $q<1$ is justified. Of course $q<1$ means that the electron spatial distribution is more concentrated near the positive ion than the distribution responsible of DH screening which is the Coulomb potential times an exponential factor.\\
Moreover, we have two coordinate spaces to consider: spatial coordinate and momentum coordinate.
If $q$ is a universal parameter for the observed system the same $q$ should be used for the spatial and for the  momentum (energy) coordinates system. We assume that the mechanism that motivates a $q<1$ distribution, due to the density fluctuation, is also responsible, through the relation density-temperature-momentum (energy) of the momentum $q<1$ distribution. Therefore also the momentum component of the distribution is a non extensive Tsallis distribution with $q<1$. On the other hand, if the same $q$ occurs for spatial coordinate and for momentum coordinate distributions, this could imply some problems with uncertainty principle with $q<1$ systems \cite{Wilk2}.\\
Other requirements and observations impose to use a $q<1$ distribution in the momentum space.\\
For instance, in the study of the spectral lines of the Sun a cutoff is required. As shown, for example in \cite{Eletskii,Chevallier} the electron component of a central core stellar plasma is made by correlated electrons where fluctuations of several types occur. The equation of state contains many non standard terms and we can realize that in such a case statistical distribution must deviate from MB one.\\
Although very recent experiments at GSI concern with oscillation of decay constants (known as GSI anomaly) we do not discuss here this case between the environment effect could be absent due to the specific experimental conditions \cite{Vetter,Litvinov,Herlert}.

In this work we want to limit ourselves to discuss only two nuclear problems of astrophysical interest where we use the Modified Debye-H\"uckel (MDH) approach. Before this discussion we describe how one can go beyond the DH approach and derive a screening potential in an astrophysical plasma using a non extensive $q<1$ electron distribution (Section 2). As an application of the MDH potential we study the d--d fusion rate in metal matrices. Also in this case we take $q<1$ and show that a good agreement with the experimental results can be obtained (Section 3). The electron capture decay by $^7$Be is discussed, the variation of the rate evaluated and few  implications reported about the neutrino fluxes (Section 4). Then we discuss the meaning, on a microscopic level, of taking $q<1$ for particle distributions. Formation of a distribution with $q<1$ is discussed and conclusions are reported (Section 5).


\section{Modified Debye-H\"uckel screening}

Thermal effects and  screening phenomena in plasma environment are strictly connected.
Assuming a MB thermal distribution, the DH ion-screening potential
\begin{equation}
V_{DH}(r)=\frac{Z_1\,Z_2\,e^2}{r}\,\exp\left(-{r\over R_{\rm DH}}\right) \ ,
\end{equation}
is obtained through the Poisson equation, where
\begin{equation}
R_{\rm DH}=\sqrt{k\,T\over4\pi\,n\,Z_\rho\,e^2} \ ,
\end{equation}
is the DH radius, but non-local thermodynamic effects can deviate the high energy tail of the velocity distribution from the exponential MB form.\\ Recently \cite{Quarati1}, it has been proposed an approach based on the application of super-statistic and/or a $q$-version of the Poisson equation to the DH screening model by assuming that non-linear effects produce fluctuations on the inverse of the DH radius $1/R_{DH}$, with a Gamma-function probability distribution
\begin{equation}
f_q(r,\,\lambda,\,\lambda_0)=\frac{A_q\,(r,\,\lambda_0)^{\frac{1}{1-q}}}{\Gamma\left(\frac{1}{1-q} \right)}
\lambda^{\frac{1}{1-q}-1}\,e^{-\lambda\,A_q(r,\,\lambda_0)} \ ,
\end{equation}
where $f_q(r,\,\lambda,\,\lambda_0)$ represents the probability density to observe a certain value $\lambda$ spreads around a central value $\lambda_0\equiv\langle\lambda\rangle$, with
\begin{equation}
\langle\lambda\rangle=\int_0^\infty f_{q}(r,\,\lambda,\,\lambda_0)\,\lambda\, d\lambda\equiv\left\langle\frac{1}{R_{DH}}\right\rangle \ .
\end{equation}
To obtain from $f_q(r,\,\lambda,\,\lambda_0)$ an electron depleted tail distribution with a cutoff,
we limit the entropic index $q$ into the $0\leq q \leq 1$ interval. Furthermore, we assume
\begin{equation}
A_q(r,\,\lambda_0)=\frac{1}{(1-q)\,g(q)\,\lambda_0}-r \ ,
\end{equation}
where $g(q)$ is a generic entropic index function that satisfies the condition $g(1)=1$
(in \cite{Quarati1,Quarati2} the choice  $g(q)=1/(2-q)$ is adopted).\\
The point charge potential $V_q(r)$ can be identified with the functional
\begin{equation}
\mathcal{F}_q(r,\,\lambda_0)=\int\limits_0\limits^\infty f_{q}(r,\,\lambda,\,\lambda_0)\,e^{-\lambda\,r}\, d\lambda \ ,
\end{equation}
through the relation
\begin{equation}
r\,V_q(r)=Z_1\,Z_2\,e^2\,\left\langle\frac{1}{R_{DH}}\right\rangle^{-1}\,\mathcal{F}_q(r,\,\lambda_0) \ .
   \label{eq:PotentialFunctionalRel}
\end{equation}
The charged particles distribution $\rho_q(r)$ and the point charge potential $V_q(r)\propto \rho_q(r)$ can be derived from Eq. (\ref{eq:PotentialFunctionalRel}), developing the functional $\mathcal{F}_q(r,\,\lambda_0)$ with the previous assumption.
A cutted form for the potential is obtained as
\begin{equation}
V_q(r)=\frac{Z_1\,Z_2\,e^2}{r}\,\exp_q
\left(-\frac{r}{\xi_q}\right) \ ,
  \label{eq:DebHuckModifPotRevised}
\end{equation}
where
\begin{equation}
\xi_q=\Big(g(q)\,\langle1/R_{\rm DH}\rangle\Big)^{-1} \ ,
\end{equation}
and
\begin{equation}
\exp_q(x)=[1+(1-q)\,x]_+^{1/(1-q)} \ ,
\end{equation}
is the $q$-exponential function with $[x]_+=x$ for $x\geq0$ and $[x]_+=0$ for $x<0$. In this way, $V_q(r)$ vanishes for $r\geq\xi_q/(1-q)$.\\
The main contribution to the charged particles fusion cross section is given by the screening barrier penetration factor $P(E)$.
In the standard DH potential case, the simplified expression
\begin{equation}
\sigma(E)=\frac{S(E)}{E}\,P(E) \ ,
\end{equation}
can be obtained ($S(E)$ is the astrophysics factor), that differs from the bare nuclei cross section $\sigma_{\rm bare}=S(E)/E$ only for the penetration factor
\begin{equation}
P(E)=\exp\left(-\pi\,\sqrt{\frac{E_{\rm G}}{E+U_{DH}}} \right) \ ,
\end{equation}
where
\begin{equation}
U_{DH}=\frac{D}{R_{DH}} \ ,
\end{equation}
with $D=Z_1\,Z_2\,e^2$ and $E_{\rm G}$ is the Gamow energy.\\
Differently, in the case of the MDH potential $V_q(r)$, the penetration factor $P_q(E)$ is given by the expression:
\begin{equation}
P_q(E)=\exp\left(-\frac{2}{\hbar\,c}\int\limits_0\limits^{r_q}\sqrt{2\,\mu\,c^2\,(V_q(r)-E)}\,dr \right) \ ,\label{eq:PenBarrierDHmodifScreen}
\end{equation}
where the classical turning point $r_q$ has to be determined through the equation $V_q(r_q)=E$.\\
Equation (\ref{eq:PenBarrierDHmodifScreen}) can be analytically solved in the $q=0$ case, obtaining
\begin{equation}
r_q={D\over E+g(0)D} \ ,
\end{equation}
and
\begin{equation}
P_0(E)=\exp\left(-\pi\,\sqrt{\frac{E_{\rm G}}{E+g(0)\,D}}\right) \ .
\end{equation}
The bare nuclei cross section can be corrected, to account for the MDH screening, multiplying $\sigma_{\rm bare}(E)$ by the factor
\begin{equation}
P_q(E)\,\exp\left(\pi\,\sqrt{\frac{E_{\rm G}}{E}}\right) \ .\label{eq:ModifDHpotCorrFact}
\end{equation}
In conclusion the penetrating charge gains the energy $U_{\rm DH}$ in the DH approach and the energy $g(0)\,D$ in the MDH approach when $q=0$.\\
Let us remark that the function $g(q)$ is not arbitrary but is related to the charge distribution $\rho_q$. In fact, it is easy to show that for $q<1$
\begin{eqnarray}
\rho_q(r)\sim-{1\over\xi_q\,r}\,\exp_q\left(-{r\over\xi_q}\right) \ ,
\end{eqnarray}
so that the screening charge is distributed
from $r=0$ to $r_{\rm cut}=1/[(1-q)\,\xi_q]$.
Therefore if we have information from experiments or from models on the charge distribution we can fix the function  $g(q)$ and its value. For instance, in \cite{Saltzmann} the distribution around deuterons is evaluated by means of a Thomas-Fermi model.


\sect{Anomalous enhancement in low energy d-d fusion rate}

In recent years a number of different experiments, with target  adsorbed in a metallic matrix, have
evidenced a strong enhancement in the fusion reaction rate at few $keV$ \cite{Assenbaum,Raiola1,Raiola2,Coraddu,Saltzmann}. For instance, the d(d, t)p reaction has been widely investigated and $^{6,7}$Li(d, $\alpha)^{4,5}$He has been studied with similar results.\\
A less strong enhancement has been observed in gas target experiments, that can be easily explained by the standard electron screening, with a potential $U_{\rm e}$\/ of the same order of the adiabatic limit $U_{\rm ad}=28\,eV$.\\
However in deuterated metal target experiments, a potential $U_{\rm e}$ of hundreds of $eV$, ten times greater than the limit $U_{\rm ad}$, is needed to reproduce the  results.\\
A possible explanation has been proposed (see \cite{Assenbaum,Raiola1,Raiola2,Coraddu,Saltzmann}) based on a simplified model of the classical quasi-free electron. It predicts an electron screening distance of the order of the Debye length.\\
This approach reproduces both the correct size of the screening potential $U_e$ and its dependence on the temperature: $U_{\rm e}\propto T^{1/2}$, but the mean number of quasi-free
particles in the Debye sphere results much smaller than one. Then, the picture of the Debye screening, which should be a cooperative effect with many participating particles,  seems not to be applicable.\\
The thermal motion of the target atoms is another mechanism capable of increasing the reaction rate; however, Maxwellian momentum distribution at the experimental temperatures gives negligible effects. The relationship between energy and momentum of quasi particles can be  broaden by  many-body collisions, then a long tail, non-Maxwellian momentum distribution can emerge from a MB energy distribution. Fusion processes select high-momentum particles that are able to penetrate the Coulomb barrier and are, therefore, extremely sensitive probes of the distribution
tail \cite{Starostin,Coraddu1}. This quantum dispersion effect has been already introduced as a possible explanation of the reaction rate enhancement. The screening plasma particle effect on the reaction rate has been  evaluated adopting for the first time, the MDH potential proposed in \cite{Quarati1}.\\
For instance we compare the MDH astrophysical factor for d-d reaction with the
experimental data reported in \cite{Raiola1,Raiola2}, adopting the choice $g(E)=3-2q$.
\begin{figure}[ht!]
\begin{center}
\includegraphics*[height=6cm,width=8.4cm]{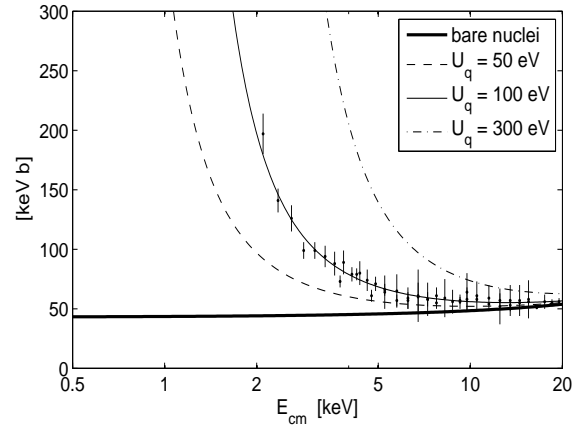}
\caption{\footnotesize Astrophysical Factor experimental points from ref. \cite{Raiola1,Raiola2}.
Bare nuclei curve correspond to $S_{\rm bare}=43+0.54\,E_{\rm cm}$ keV b,
while the screened curves are $S=f_q\cdot S_{\rm bare}(E_{\rm cm})$, $U_q=e^2\,\langle1/R_{DH}\rangle$, $g(q)=3-2q$ and $q=0$. Ion thermal motion is neglected: $E_{\rm cm}=E/2$.}
\label{fig:ModifDHRaiola2002Exp}
\end{center}
\end{figure}
The results is shown in Fig. \ref{fig:ModifDHRaiola2002Exp}
for the entropic index $q=0$.\\ One can observe as a screening potential $U_q$ three times lower than in the  standard DH case ($U_q\sim100\,eV$ instead of $U_{DH}\sim300eV$), is required to reproduce the experimental data. An electrostatic screening potential of this order of magnitude has been obtained, for instance, by Saltzmann and Hass \cite{Saltzmann} through a Thomas-Fermi model of the electron gas in a deuterated-copper target
(they obtained a screening potential of $163\,eV$, instead of the $470\,eV$ needed to reproduce the experimental results).\\ Now we are treating $g(q)$ and $q$ as free parameters, but, in principle, a link can be establish between the inverse DH radius, the temperature fluctuations and the $q$-index:
\begin{equation}
\frac{\Delta\left(1/R_{\rm DH}\right)}{1/R_{\rm DH}}
=\frac{\Delta(k\,T)}{k\,T}=\sqrt{1-q} \ .
\end{equation}
By this way the modified DH potential can be obtained starting from the environment condition.\\
We conclude: the reaction cross section enhancement, due to plasma screening effects, has been evaluated adopting the MDH potential. As shown in \cite{Assenbaum,Raiola1}, accounting the matrix metal valence electrons, a screening potential $U\approx 100\,eV$ can be obtained, greater than the adiabatic limit $U_{\rm ad}$ but even too low to reproduce the experimental results. We showed as this discrepancy can be removed adopting the MDH potential (\ref{eq:DebHuckModifPotRevised}) with a proper choice for the function $g(q)$ and the entropic index $q$, as can be seen in Fig. \ref{fig:ModifDHRaiola2002Exp}. In principle the value of the entropic index $q$ can be derived from the plasma equation of state, then the MDH screening contribution to the rate enhancement can be evaluated exactly. We will investigate this last point in a future work.


\sect{Electron capture by $^7$Be}

Electron capture (EC) is a nuclear process where an internal proton of a nucleus is converted to a neutron by means of  weak interaction: e$^-+$(Z$+1,\,$A)$\to$(Z$,\,$A)$+\nu$.\\
In laboratory, this reaction proceeds by capture of a K-shell electron since these electrons have the greatest probability to move in the nuclear region. In the stellar core, atoms are almost completely ionised so that EC proceeds with continuum electrons and the rate depends on the electron density over the nucleus. In plasmas of very high density, stable nuclei are forced to capture free energetic electrons and hydrogen gas change in a neutron gas.\\
In stars like Sun, the capture e$^-+^7$Be$\to^7$Li$+\nu$ is one of the most important $\beta$-decay reaction. In a star in nuclear equilibrium, $\beta$-decay plays a special role because the inverse reaction requires the absorption of a neutrino. The neutrino escapes from the local environment with an average energy $E_\nu=0.814\,MeV$.
Therefore, $\beta$-decays cannot participate to a true equilibrium; they are sufficiently slow so that an equilibrium can be established in a time short compared to the time required for a significant change of the element average composition. However, environments must be such to satisfy this time condition and this can be so if only frequent and pure Coulomb interactions are active.\\
The principle of detailed balance ensures that in global thermodynamic equilibrium loss of electrons captured by nuclei in the medium-high energy tail is counterbalanced by production of such electrons due to inverse reaction. However, neutrinos, once produced, are travelling away with their kinetic energy and the intermediate energy region of electron distribution results depleted compared to Maxwellian distribution. In fact, the main contribution to the EC rate is from the region at $2--3$ time $k\,T=1.27\,keV$. Electron component is a non-LTE state. This metastable state can be well described by a $q$-distribution function with $q<1$.

It is known that in nuclear continuum EC the rate includes the Fermi factor, i.e. the electronic density at the limit of $r\to0$ in a pure Coulomb potential. The pure nuclear cross section is corrected by a factor because of the Coulomb interaction between the captured electron and the nucleus. Usually, in a plasma environment, DH potential takes the place of Coulomb potential. In this case the electron density at the nucleus is known only numerically from the solution of the appropriate Schr\"oedinger equation.\\
In an astrophysical plasma environment, the momentum distribution of screening electrons can differ significatively from MB distribution. Here we use the MDH potential described previously.\\
Of course, also in this case we need to evaluate the electron density at $r=0$ by solving the appropriate Schr\"odinger problem for electrons in the continuum.\\
Whereas the solution with a Hulth\'{e}n potential can be given in a close analytic form,
the MDH potential (as well as the standard DH potential) admits only numerical solutions for the electron density at $r=0$. Alternatively, one can use the Hulth\'{e}n potential that fits quite well the MDH potential in the small $r$ region (near the nucleus), but contains an infinite tail instead of a cut-off at an appropriate value of $r\equiv r_{\rm cut}$, as for the MDH potential with $q<1$.\\
In the following, we evaluate the rate for the free electron capture by a (A$,\,$Z)-nucleus, given by the integral, in the three dimensional space of velocities, of electron capture cross-section $\sigma_{\rm e}$ times the electron velocity $v$, the normalised probability density $(F_{_{\rm C}},\,F_{_{\rm DH}},\,F_{_{\rm H}}$ or $F_q)$\footnote{The normalised probability density that an electron of the continuum spectrum with velocity $v$ and travelling in a screening potential $(V_{\rm C},\,V_{\rm DH},\,V_{\rm H}$ or $V_q$) be at the nucleus with coordinate $r=0$} and the normalised velocity distribution for electrons given by the function
\begin{equation}
f_q(v)=B_q\left({m_{\rm e}\over2\,\pi\,k\,T}\right)^{3/2}\,\exp_q\left(-{m_{\rm e}\,v^2\over2,k\,T}\right) \ ,
\end{equation}
where $q=1$ for C, DH and H (with $f_{_{q=1}}(v)\equiv f_{_{\rm MB}}(v)$ the normalized MB distribution).\footnote{For details on this approach, expressions used and complete numerical results see reference \cite{Quarati2}.}
We define the pure Coulomb nuclear electron capture rate, averaged over a MB distribution, as
\begin{equation}
{\cal R}_{_{\,\,\,\rm C}}(T)=\int\limits_0\limits^\infty(\sigma_{\rm e}\,v)\,F_{_{\rm C}}(E)\,f_{_{\rm MB}}(v)\,4\,\pi\,v^2\,dv ,\label{rc}
\end{equation}
where
\begin{equation}
\sigma_{\rm e}={G^2\over\pi\,(\hbar\,c)^4}\,{c\over v}\,\Big(W_0+W\Big)^2\,\chi \ ,
\end{equation}
is the nuclear electron capture cross section \cite{Quarati2}, with $G$ the Fermi constant,
$W_0$ the nuclear energy release for one electron with total energy $W$, $\chi=C_{\rm V}^2\,\langle1\rangle^2+C_{\rm A}^2\,\langle\sigma\rangle^2$
the well-known reduced nuclear matrix element.\\
The Fermi factor for Coulomb potential, given by
\begin{equation}
F_{_{\rm C}}(E)={2\,\pi\,\eta\over1-e^{-2\,\pi\,\eta}} \ ,
\end{equation}
with $\eta=4/(a_0\,p)$, where $a_0$ is the Bohr radius and  $p=m_e\,v$ is the electron momentum, follows from the definition
\begin{equation}
F_{_{\rm C}}(E)=\lim_{r\to0}\Big|{\psi_{_{\rm C}}(r)\over p\,r}\Big|^2 \ ,
\end{equation}
where $\psi_{_{\rm C}}(r)$ is the wave function of the Schr\"odinger equation with the Coulomb potential.\\
The non extensive rate ${\cal R}_{\,\,\,q}$, can be obtained by substituting in Eq. (\ref{rc}) the factor
$F_{_{\rm C}}(E)$ with the Fermi factor
\begin{equation}
F_q(E)=\lim_{r\to0}\Big|{\psi_q(r)\over p\,r}\Big|^2 \ ,
\end{equation}
where $\psi_q(r)$ can be obtained as a numerical solution of the Schr\"oedinger equation with the MDH potential \cite{Quarati1}
\begin{equation}
V_q(r)=-{Z\,e^2\over r}\,\exp_q\left(-{r\over\xi_q}\right) \ ,
\end{equation}
and $g(q)=1/(2-q)$.\\
Consistently with the derivation of the MDH potential in the definition of ${\cal R}_{\,\,\,q}$ we must insert in place of the MB distribution the normalised non extensive distribution $f_q(v)$.
The integral for the rate ${\cal R}_{\,\,q}$, when $q<1$, is performed over the real interval $[0,\,v_{\rm cut}]$ with
\begin{equation}
v_{\rm cut}=\sqrt{2\,k\,T\over(1-q)\,m_{\rm e}}<1 \ ,
\end{equation}
which defines a cut-off condition in the velocity space. When $q\to1$ the rate ${\cal R}_{\,\,\,q}$ reduces to DH rate ${\cal R}_{_{\,\,\,\rm DH}}$.

Although electron density at $r=0$ due to $V_q(r)$ is smaller than Coulomb density, in the velocity space the probability density in the low momentum region is greater than MB because the continuum electron distribution $f_q(v)$ we use privileges low momentum electrons. Therefore, screening may be important in continuum EC rate. We have calculated deviations of the rate ${\cal R}_{_{\,\,\,\rm X}}(T)$ (with X = DH, H and $q$) respect to ${\cal R}_{_{\,\,\,\rm C}}(T)$, at $k\,T=1.27\,keV$ (where EC by $^7$Be takes place).\\
For any $q<1$, ${\cal R}_{\,\,\,q}>{\cal R}_{_{\,\,\,\rm C}}$. We have verified that deviations depend very smoothly on $k\,T$ except for $\xi\leq0.45\,a_0$ and depend very strongly on $q$.\\
The value of $q$ for EC by $^7$Be in solar plasma can be derived from the expression that links $q$  to fluctuation of $1/R_{_{\,\,\,\rm DH}}$ \cite{Quarati2}.
In the solar core, where the average electron density is $n_{\rm e}=9.1\,a_0^{-3}$ and the number of particles inside the Debye sphere $N_{_{\rm DH}}$ is about $4$, we can obtain $q=0.86$. It is more safe to consider a range of values of $q$ between $0.84$ and $0.88$. At $k\,T=1.27\,keV$ and $\xi_q=0.45\,a_0$ the calculated ${\cal R}_{\,\,\,q}(T)$
is estimated to be about $7$ -- $10\,\%$ larger than standard DH ($q=1$) estimate that is the $0.69\,\%$ smaller, at the same conditions, than Coulomb rate ${\cal R}_{_{\,\,\,\rm C}}(T)$. Of course, a smaller value of $q$ should imply a much greater enhancement of EC rate over DH one.\\
Let us consider the $^7$Be -- p fusion. This reaction producing $^8$B and, as a consequence, $^8$B neutrinos, in competition with $^7$Be electron capture. We have verified that the effect of the MDH potential over its rate is negligible. In fact, correction to $F_{_{\,\,\,\rm C}}$ is effective only at relative $^7$Be -- p energies lower than $2.4\,keV$ where fusion cross section has a negligible value because its most effective energy is at $18\,keV$. Therefore, if EC rate of $^7$Be increases over its standard evaluation of a given percentage, $^7$Be increases its destruction while the neutrino flux from $^7$Be does not change because the $^7$Be density decreases. However, the $^8$B flux should diminish of the same percentage. This behaviour is in line with what is found in experiments.


\sect{Problems about distribution with $q<1$ its formation and conclusions}

Among the many astrophysical problems and experimental observations listed in the Section 1, reporting   deviations from the standard behaviour detected and measured\footnote{for some of them results should be confirmed.} that, to our opinion, showing how the environment effects could be possibly taken into account by means of $q$-thermostatistics and/or superstatistics approaches, we have selected two examples: the d--d fusion in metal matrix and the solar EC by $^7$Be.\\
Measured rates of the first subject are strongly enhanced respect to standard laboratory measures. We have proposed an explanation based on the $q=0$ thermostatistic description of the MDH potential used for the d--d fusion reactions. The parameter $q=0$ means that a great deformation of the standard MB momentum distribution function should result evident.\\
In the second subject, the electrons that screen the capture have a spatial distribution with $q=0.84-0.88$. Although in this case deformation is small, consequences in the production of $^8$B solar neutrinos are of great interest.\\
In both cases, the parameter $q$ is lower than $1$ and in the second subject the $q$-approach is used for the derivation of the MDH potential in the coordinate space and for the calculation of the rate in the momentum coordinate space and the same value of $q$ is assumed.\\
We have tried to show how the case $q<1$ be important in many nuclear physics problems in a non extensive  $q$-environment. Of course also the case $q>1$  has important applications when, for example,
gamma rays produced in the plasma comptonise electrons making a fat distribution tail.\\
However, to handle with the $q<1$ case means that we have to pose attention to some problems that we list below:
\begin{enumerate}
\item When using superstatistics to derive $q<1$ distribution. Because the average value depends on a variable, a second average is needed;
\item The function $g(q)$ of expression $1/(2-q)$ or $3\,q-2$ (see the derivation of MDH), which goes to $1$ for $q$ going to $1$, is arbitrary and we must find a condition for its determination. This can be accomplished if we know the electron charge distribution around the ion, as we have already discussed in Section 2;
\item When the process requires a double use of non-extensivity (we have the system in phase space) in the spatial coordinate and in the momentum coordinate, we must establish rules in order to know if the same $q$ should be adopted or two different values and if they are possibly linked in some way, i.e. if an uncertainty principle holds;
\item Particular experimental results are modulated in time; in this case one needs a function $q(t)$ rather a parameter $q$.
\end{enumerate}
We want now to show how  $q<1$ distribution can be formed starting from MB distribution. \\
The $q<1$ distribution has a depleted tail with a cut-off and an enhanced head. To understand how this distribution can be formed, let us start by considering a system of $N$ noninteracting particles distributed along a Maxwellian shape, at temperature $T$.\\ We introduce a quantity called cut-off energy $\varepsilon_{\rm cut}$, analogous to the Fermi energy level of quantum distributions, located at $\varepsilon_{\rm cut}=k\,T/(1-q)$. Depending on the value of $q$, with $0\leq q\leq1$, the cut-off energy may assume a value in the range $k\,T\leq\varepsilon_{\rm cut}\leq\infty$.\\
The number of particles $N_+$ in the Maxwellian distribution with an energy $\epsilon$ above the cut-off energy, i.e. within $\varepsilon_{\rm cut}\leq\epsilon\leq\infty$, can be easily calculated by
\begin{equation}
N_+=N\,{2\over\sqrt{\pi}}\,\Gamma\left({3\over2},\,{1\over 1-q}\right) \ ,
\end{equation}
where $\Gamma(a,\,x)$ is the inverse incomplete gamma function.\\
We assume that at a certain time a many-body interaction be active among the particles. As a consequence of {\em complete occupancy principle}, all the particles occupying the states above $\varepsilon_{\rm cut}$ displace below leaving all states above empty. Then, particles reorganize collectively their distribution at the appropriate temperature $T^\prime< T$, where $T^\prime={2\over5-3\,q}\,T$ spending the work ${\cal L}^{^{\rm R}}_{q<1}$.\\
The complete energy balance of this ideal process producing a depleted distribution is given by
\begin{eqnarray}
\nonumber{3\over2}\,N\,k\,T^\prime&=&{3\over2}\,N\,k\,T-{2\over\sqrt{\pi}}\Big(N_++\delta N_+\Big)
\,k\,T\,\Gamma\left({5\over2},\,{1\over1-q}\right)\\
\nonumber
& &+{2\over\sqrt{\pi}}\Big(N_++\delta N_+\Big)
\,k\,T\,\gamma\left({5\over2},\,{1\over1-q}\right)+{\cal L}^{^{\rm R}}_{q<1} \ ,\\
\end{eqnarray}
where $\gamma(a,\,x)=\Gamma(a)-\Gamma(a,\,x)$ is the incomplete gamma function.
Since generally $\delta N_+/N$ is negligible, we can write
\begin{eqnarray}
{\cal L}^{^{\rm R}}_{q<1}&\approx&-{3\over2}\,N\,k\,T\,\left[3\,{1-q\over5-3\,q}+Q_q\right] \ ,
\end{eqnarray}
where
\begin{equation}
Q_q={2\over\sqrt\pi}\,\left[1-{8\over3\,\sqrt\pi}\,\Gamma\left({5\over2},\,{1\over1-q}\right)\right]
\,\Gamma\left({3\over2},\,{1\over1-q}\right) \ .
\end{equation}
We remark that ${\cal L}^{^{\rm R}}_{q<1}/N\,k\,T$ is always negative because the work is done in favor of the environment. It depends on $q$ only and shows a maximum at $q\approx0.7$. \\
We note that MB distribution of $N$ non interacting particles in thermodynamical equilibrium can be viewed  as a generalized distribution with $q<1$ and $N_+$ quasi-particles below $\varepsilon_{\rm cut}$ disposed above if the energy level sequence remans unchanged.\\
We remember that $q<1$ case is related to a finite heath bath as shown in the past by Plastino and Plastino \cite{Plastino,Vignat1,Vignat2}.\\
In recent years, many experiments have shown results that deviate from the estimations evaluated on the basis of global thermodynamic equilibrium, of independence of single events neglecting particle correlations, fluctuations and non linear effects. Under these experimental evidences we realize that many processes that were understood on the basis of simplifications can be better understood if we improve earlier evaluations by generalising previous approaches. In such situation we are convinced that $q$-thermostatistics will play a fundamental role.





\vspace{10mm}

\noindent Electronic addresses:\\
$^\ast$massimo.coraddu@ca.infn.it\\
$^\dag$marcello.lissia@ca.infn.it\\
$^\ddag$piero.quarati@polito.it\\
$^\S$antonio.scarfone@polito.it

\end{document}